# Empirical Comparisons of CNN with Other Learning Algorithms for Text Classification in Legal Document Review

Robert Keeling, Esq.
Complex Commercial Litigation
Sidley Austin LLP
Washington DC, USA
rkeeling@sidley.com

Rishi Chhatwal, Esq.
Legal
AT&T Services, Inc.
Washington DC, USA
rishi.chhatwal@att.com

Nathaniel Huber-Fliflet
Data & Technology
Ankura
Washington DC, USA
nathaniel.huber-fliflet@ankura.com

Jianping Zhang
Data & Technology
Ankura
Washington DC, USA
jianping.zhang@ankura.com

Fusheng Wei
Data & Technology
Ankura
Washington DC, USA
fusheng.wei@ankura.com

Haozhen Zhao
Data & Technology
Ankura
Washington DC, USA
haozhen.zhao@ankura.com

Shi Ye
Data & Technology
Ankura
Washington DC, USA
shi.ye@ankura.com

Han Qin
Data & Technology
Ankura
Washington DC, USA
han.qin@ankura.com

*Abstract*— **Research has shown that Convolutional Neural Networks (CNN) can be effectively applied to text classification as part of a predictive coding protocol. That said, most research to date has been conducted on data sets with short documents that do not reflect the variety of documents in real world document reviews. Using data from four actual reviews with documents of varying lengths, we compared CNN with other popular machine learning algorithms for text classification, including Logistic Regression, Support Vector Machine, and Random Forest. For each data set, classification models were trained with different training sample sizes using different learning algorithms. These models were then evaluated using a large randomly sampled test set of documents, and the results were compared using precision and recall curves. Our study demonstrates that CNN performed well, but that there was no single algorithm that performed the best across the combination of data sets and training sample sizes. These results will help advance research into the legal profession's use of machine learning algorithms that maximize performance.**

## I. Introduction

Due to the rapidly growing volume of electronically stored information, the costs involved in manually reviewing documents in the legal industry have grown dramatically. Companies regularly spend millions of dollars responding to document requests [1]. To more efficiently cull through massive volumes of data for relevant information, attorneys have begun using text classification, a supervised machine learning technique typically referred to as predictive coding or technology assisted review (TAR).

Logistic Regression (LR) and Support Vector Machine (SVM) are two popular machine learning algorithms used in predictive coding [2]. Recent research has shown that Convolutional Neural Networks (CNN) with word embedding can also be effectively applied to text classification [4,5,6,7]. But most of these studies were conducted on data sets with short documents containing a small number of words. In actual legal document reviews, document lengths will vary from a few sentences to a few hundred pages of words. In a 2018 study [3], we applied CNN to different data sets from actual legal matters containing documents of varying word lengths and compared the results of CNN with SVM. Our study found that while CNN obtained better precision and recall metrics when using large training sets, it did not perform as well as SVM when using smaller training sets.

In this paper, we report our further research into the performance of CNN as part of a predictive coding protocol with a more comprehensive analysis that tuned the hyperparameters of CNN and compared it with three popular machine learning algorithms: Support Vector Machine, Logistic Regression, and Random Forest. For the three learning algorithms, we also tuned their hyperparameters to optimize performance. The CNN model we developed for this study was a single layer convolution with max pooling that was based on prior research into CNN [4, 5]. While other research into CNN has used more complex algorithms [9,10], these studies tend to require larger training sets and longer training time. For document reviews in the legal industry, a simple model is a practical choice that allows for faster training and predicting time.

We begin by describing the settings of the four machine learning algorithms in Section II. Data sets and experimental design are described in Section III. We report our results in Section IV and the paper concludes with Section V.

## II. SETUP OF LEARNING ALGORITHMS

### A. Convolutional Neural Network (CNN)

We used the simple CNN model introduced in a 2014 study [4] that consists of a single one-dimension convolution layer followed by dropout, one-dimension max pooling, and a fully connected layer with binary classification. We chose to use an embedding layer as part of training instead of the pretrained word embedding used in the 2014 study [4]. This is the same structure used in our own 2018 study [3], however, we modified several hyperparameters:

- Filter Number: 64
- Filter Kernel Size: 2
- Maximum Pooling Size: max (i.e. 1-max)
- Tokenizer Vocabulary Size: 20,000
- Tokenizer Sequence Length: 2000

These parameters were chosen based on experiments with the legal matter data sets that we used in this study. In prior research [4,5,6], authors have discussed various hyperparameter settings, and our parameter choices followed some of their recommendations, including using 1-max pooling. These hyperparameters settings were fixed across data sets and training set sizes. But the dropout rate was customized for each training set – the dropout rate and epochs parameters were chosen by analyzing the results of a grid search of various combinations and identifying the combination with the optimal precision rate at 75% recall.

We used Keras with TensorFlow backend to implement the convolution neural network. The summary of the parameters we used for the convolution neural network is shown in Table 1.

*Table 1. Keras Model Summary*

```
Layer (type)                    Output Shape           Param #
embedding_1 (Embedding)         (None, 2000, 100)      2000000
dropout_1 (Dropout)             (None, 2000, 100)      0
conv1d_1 (Conv1D)               (None, 1999, 64)       12864
max_pooling1d_1 (MaxPooling1    (None, 1, 64)          0
flatten_1 (Flatten)             (None, 64)             0
dense_1 (Dense)                 (None, 1)              65
```

### B. Support Vector Machine (SVM), Logistic Regression (LR), and Random Forest (RF)

We used the scikit-learn packages for traditional machine learning algorithms. For SVM, we use LinearSVC with grid search cross validation on the penalty parameter c of the error term; for LR, we also used grid search cross validation on c and classifier solvers (between liblinear and newton-cg). For RF, we used 300 trees.

## III. DATA SETS AND PREPROCESSING PARAMETERS

### A. Data Sets

We conducted experiments on four data sets, named A, B, C, and D, from confidential, non-public, real legal matters across various industries such as social media, communications, education, and security. Attorneys reviewed all documents in the four data sets over the course of the legal matters. Table 2 shows the statistics of the four data sets.

*Table 2: Data Set Statistics*

| Data Set | Total Documents | Responsive Documents | Not Responsive Documents | % Responsive |
|---|---|---|---|---|
| A | 410,954 | 81,324 | 329,630 | 19.8% |
| B | 1,570,956 | 408,897 | 1,162,059 | 26.0% |
| C | 492,318 | 201,147 | 291,171 | 40.1% |
| D | 308,738 | 46,644 | 262,094 | 15.1% |

For each data set, we randomly selected 100,000 labeled documents as the corpus for our experiments. Out of each of set of 100,000 labeled documents, we set aside 10,000 randomly selected documents as the test set. Then we generated four incremental training sets of sizes 2,500, 5,000, 10,000 and 25,000, respectively, by randomly selecting them from the remaining labeled documents. We applied up-sampling for data set D to increase the responsive label ratio to 50% for the training sets, while keeping the ratio unchanged for the other three data sets.

### B. Text Preprocessing

For CNN, we used Keras sequence model tokenizer to prepare inputs for the CNN algorithm from the training sets with the specified vocabulary size and sequence length. The same text preprocessing function [2] was used for the SVM, LR, and FR algorithms. The text preprocessing parameters we used consisted of the following steps:

1. Tokenization,
2. Token Filtering,
3. Stemming,
4. N-gram Generation, and
5. Feature Selection.

*Tokenization* breaks up the sequence of strings (sentences) in a document into a set of smaller units, such as words, called tokens. *Token Filtering* removes irrelevant tokens such as stop words, numbers, short words (e.g., words with one or two characters), and long words (e.g., words with more than 20 characters). *Stemming* converts words into their root forms. The *N-gram Generation* step generates all n-grams of a document as features to represent the document. The *Feature Selection* step applies a feature selection algorithm to the training documents to identify a subset of the most effective features (words or n-grams) to represent a document.

In these experiments, we removed stop words and numbers; no stemming was applied; 1-gram was used; normalized frequency was used; and 20,000 tokens were selected as features.

## IV. RESULTS

### A. The Metrics

In predictive coding, a precision and recall curve is commonly used to evaluate performance. Practically, we evaluated precision at 75% recall. We show the results in two ways. One as a typical precision / recall curve (Figures 1-4) and another as the precision rate at a specific recall rate of 75%, a commonly used performance metric in the legal domain (Figures 5-8).

### B. Precision / Recall Curve Comparisons

For each data set, we placed the precision / recall curve for each of the four algorithms together in a single plot and then group the plots by different training set sizes (Figures 1-4). It is clear from these figures that no one algorithm significantly outperformed another across all four data sets for each of the training set sizes. Random Forest did not perform as well as other algorithms for low recalls with training set sizes of 2,500 and 5,000 documents. Although the performance of CNN can outperform the other algorithms in some cases, like the results of other research show, the performance of CNN is not significantly better than the performance of the other algorithms across all four data sets and training data sizes. Also, CNN's performance did not significantly increase with the increase of the training set size.

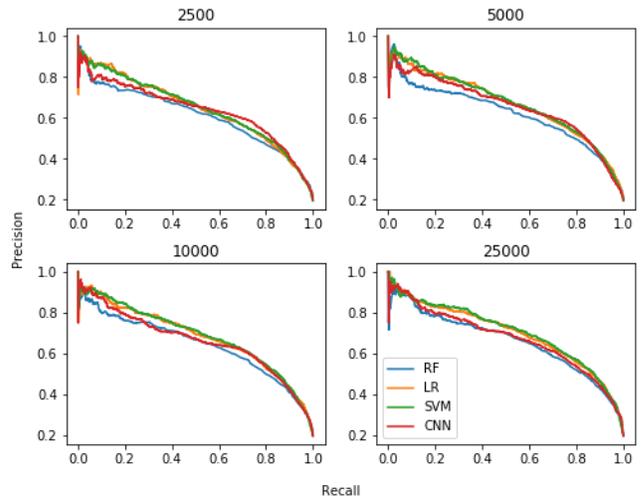

*Figure 1. Precision Recall / Curves - Data Set A*

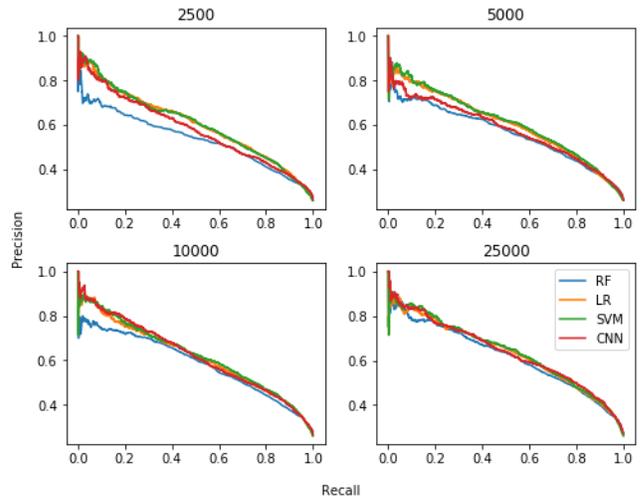

*Figure 2. Precision Recall / Curves - Data Set B*

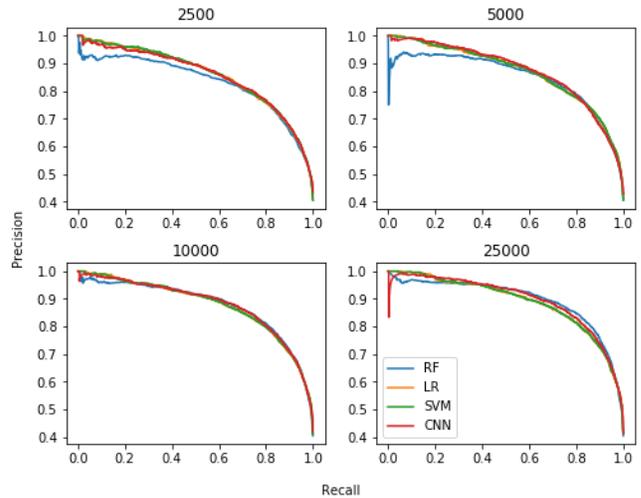

*Figure 3. Precision Recall / Curves - Data Set C*

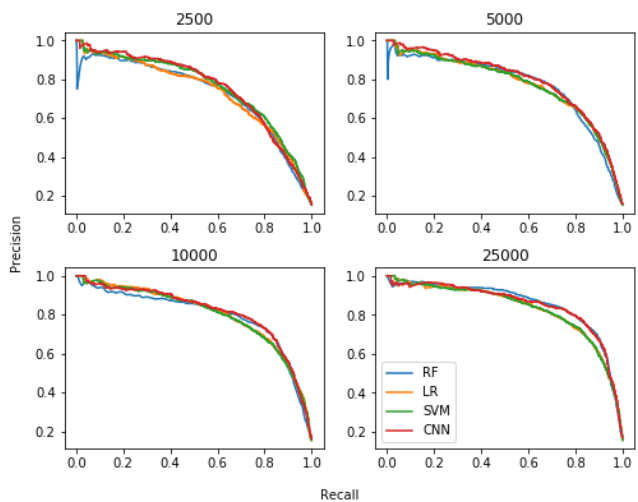

*Figure 4. Precision Recall / Curves - Data Set D*

### C. Comparisons of Precision at 75% Recall Rate

At a recall rate of 75%, Figures 5-8 show the precision rate for each data set for the four different training set sizes. Again, no algorithm performed better in all cases, but CNN achieved the highest precision among the four algorithms: nine times out of 16 experiments. See Table 3. Random Forest did not perform well on Data Sets A and B, but did perform well on Data Sets C and D. SVM and Logistic Regression performed similarly and their results are comparable with the other two algorithms in most experiments.

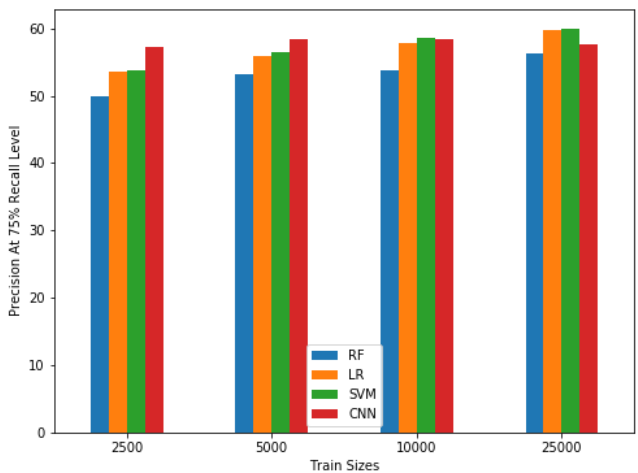

*Figure 5. Precision at 75% Recall Rate - Dataset A*

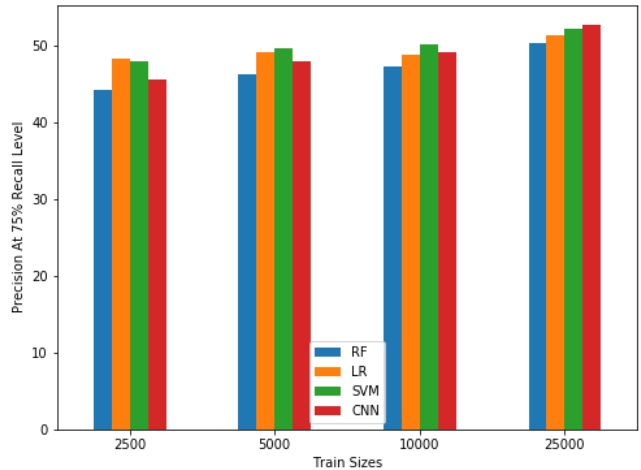

*Figure 6. Precision at 75% Recall Rate - Dataset B*

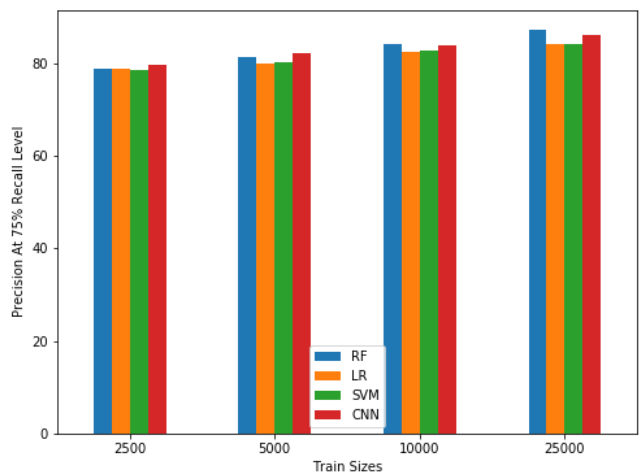

*Figure 7. Precision at 75% Recall Rate - Dataset C*

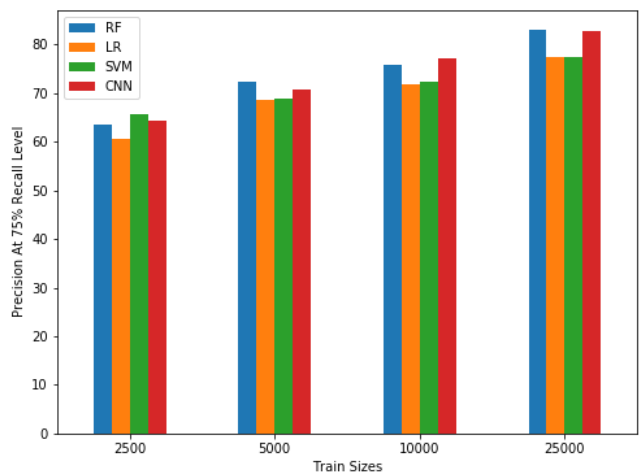

*Figure 8. Precision at 75% Recall Rate - Dataset D*

*Table 3. Top Algorithm Performers*

|   | 2500 | 5000 | 10000 | 25000 |
|---|------|------|-------|-------|
| A | CNN | CNN | SVM/CNN | SVM/LR |
| B | LR/SVM | SVM/LR | SVM | CNN |
| C | CNN | CNN | RF/CNN | RF |
| D | SVM | RF | CNN | RF/CNN |

## V.  CONCLUSION

This study demonstrates that CNN can provide an effective algorithm choice for a predictive coding process. Our results, for example, demonstrate that CNN can perform well even when using a small training set size, a different conclusion compared to our prior findings in 2018 [3]. At the same time, this study demonstrates that while some algorithms performed better than others for specific combinations, no one algorithm outperformed another across all the different combinations of data sets and training set sizes. For example, the precision rate at 75% recall experiments show that CNN performs slightly better than others on average.

REFERENCES

[1] *Nicholas M. Pace & Laura Zakaras, "Where the Money Goes: Understanding Litigant Expenditures for Producing Electronic Discovery," RAND at 17 (2012).*

[2] *Chhatwal, R., Huber-Fliflet, N., Keeling, R., Zhang, J. and Zhao, H. (2016). Empirical Evaluations of Preprocessing Parameters' Impact on Predictive Coding's Effectiveness. In Proceedings of 2016 IEEE International Big Data Conference*

[3] *Wei, F., Han, Q., Ye, S., Zhao, H. "Empirical Study of Deep Learning for Text Classification in Legal Document Review" 2018 IEEE International Big Data Conference*

[4] *Yoon Kim. 2014. Convolutional neural networks for sentence classification. arXiv preprint arXiv:1408.5882*

[5] *Rie Johnson and Tong Zhang. 2014. Effective use of word order for text categorization with convolutional neural networks. arXiv preprint arXiv:1412.1058.Yoav Goldberg. 2015. A primer on neural network models for natural language processing. arXiv preprint arXiv:1510.00726*

[6] *Y. Zhang and B. C. Wallace, "A sensitivity analysis of (and practitioners' guide to) convolutional neural networks for sentence classification," CoRR, vol. abs/1510.03820, 2015. [Online]. Available: http://arxiv.org/abs/1510.03820*

[7] *Yoav Goldberg. 2015. A primer on neural network models for natural language processing. arXiv preprint arXiv:1510.00726*

[8] *Rie Johnson and Tong Zhang. 2016. Convolutional neural networks for text categorization: Shallow word-level vs. deep character-level. arXiv:1609.00718 .*

[9] *Rie Johnson and Tong Zhang. Deep pyramid convolutional neural networks for text categorization. In ACL, pages 562–570, 2017*

[10] *Alexis Conneau, Holger Schwenk, Loïc Barrault, and Yann LeCun. 2016. Very deep convolutional networks for natural language processing. arXiv:1606.01781v1*